\def\ecr{\epsilon_{\rm cr}}
\documentstyle[12pt]{article}
\title{Cosmic Strings At the Electroweak\\ Phase Transition}
\author{{Susmita Bhowmik Duari}\\
and \\
U. A. Yajnik\\
Department of Physics, Indian Institute of Technology,\\ Bombay 400076}
\date{Oct. 29 1993}
\begin{document}
\baselineskip=24pt
\vfil
\maketitle
\begin{abstract}
If cosmic strings are present at the electroweak phase transition, they can
act as seeds on which bubbles of true vacuum nucleate. We explore the nature
of such a phase transition, in particular the wall velocity and thickness
of the bubbles.
{}From the viewpoint of electroweak baryogenesis, adiabatic conditions exist
in the expanding bubble walls, and such models of baryogenesis can be
successfully applied. In the present mechanism, the nature of the
electroweak phase transition is insensitive to the other details of the
model, thus reducing the uncertainties in the estimate of net baryon
asymmetry.
\end{abstract}
\vfil

email : yajnik@cc.iitb.ernet.in
\newpage
It is considered as established \cite{ser1}
 that {\bf B+L } anomaly of the standard electroweak
theory is unsuppressed in the high temperature phase in the early universe.
Given the conditions of approximate equilibrium, this implies zero value for
net ${\bf B+L}$ number. Various possibilities still remain open for
obtaining  the
presently observed residue of {\bf $ n_{B}/ n_{\gamma} \simeq \
10^{-10 \pm 0.5}$},
but the most promising among these is that of generating {\bf B}-asymmetry
in the course of electroweak phase transition. Achieving this possibility
requires that the electroweak phase transition
should  be first order\cite{Kuzmin}.
Investigations of the electroweak temperature dependent
effective potential indicate the phase transition to be indeed first order.
 But in considering any given
particle physics model for electroweak {\bf B} generation, one needs details
of the phase transition, such as the thickness of the bubble walls
and their speed. These questions has been extensively investigated, for
instance, in \cite{Dine.et.al}, \cite{Buchmuller}.

Here we investigate the effect on the phase transition of the presence of
cosmic strings. Many extensions of the Standard Model involve new gauge
forces. Spontaneous breaking of such symmetries generically leads to
defects such as cosmic strings.
It has been shown that cosmic strings can  significantly affect the
nature of a first order phase transition. This comes about by
the nucleation of bubbles of true vacuum on the strings\cite{Yajnik}.
The bubble nucleation occurs
at a precise temperature on all seeding strings and much sooner than is
possible by spontaneous tunneling process.
In this paper we shall determine the temperature
${T_1}$ at which bubble nucleation begins on the strings, and the thickness
and the speed of the bubble walls. These parameters are determined
as a function of the Higgs mass. Top quark mass is unimportant in this
mechanism. Using these parameters, we
comment on the viability of some of the scenarios proposed for baryogenesis
at the electroweak phase transition.

We begin with a brief discussion of the string induced phase transition
\cite{Yajnik}. Let us assume that the Standard Model gauge group
${SU(3)}_{c}\times {SU(2)}_{L}\times {U(1)}_{Y}$ is the group of the residual
symmetry after the breakdown of some larger group G, {\it e.g}, an SO(10) or
supersymmetric SU(5) unification group, or a Left-Right symmetric model. The
breakdown to the Standard Model may involve more than one stage of symmetry
breakdown and involve several scalar multiplets. Let us generically designate
one of these higgs fields by $\chi$. For string induced phase transition to
occur, it will be sufficient that one such multiplet satisfies the following
conditions. Firstly, $\chi$ must be  nontrivially involved in the
semiclassical ansatz for the strings, and the strings remain
topologically stable upto, but not
necessarily through, the electroweak phase transition. And secondly,
the quantum numbers of  $\chi$ must allow interaction terms
with $\phi$,  the standard Higgs.
Let the translation invariant vacuum value $<\chi>$ be $M_{\chi}$. Since
$\chi$ is involved in the semiclassical string configuration in a sector with
nonzero winding number, $<\chi>$ is a function of the radial distance from the
string core. Now by the usual device of maintaining hierarchy, the $\phi-\chi$
coupling terms have to be fine tuned, so that $<\phi>$ remains zero in the
translation invariant vacuum. However in the vicinity of the strings, this
hierarchy arrangement will break down and $<\phi>$ also will be a function of
radial distance from the string core, and $<\phi > \ \sim \  M_{\chi}$ in the
string core. This is the key requirement for the occurance of string induced
phase transition. Further details of how this may come about is discussed in
\cite{Yajnik}.

Let $\phi_{1}$ be the trivial and $\phi_2$ the nontrivial minima
 of the temperature
dependent effective potential $ V^{T}[\phi ]$ in the electroweak theory.
Let $T_c$ be the  temperature at which
$ V^{T}[\phi_{1}]\  =\  V^{T}[\phi_{2}]$ .
We shall define below a dimensionless parameter
 $\epsilon \ \propto \ ( T - T_{c})$. We can prove that
below a  small negative value $\epsilon_{\rm cr}$ of
$\epsilon$, there is no solution of the effective action that
is time translation invariant and approaches $\phi_1$ large
distance away from the string.
 If such a solution is given as an initial condition
for $\epsilon>0$, it is rendered unstable for $\epsilon<\ecr$.
One finds time dependence setting
in and obtains solutions representing expanding bubbles of true vacuum.
{}From numerical solutions one can determine $\epsilon_{\rm cr}$,
and the velocity of expanding bubbles under ideal conditions, {\it i.e},
ignoring the out of equilibrium interaction with the plasma. We here
determine these physical parameters as functions of the Higgs mass.
In order to do this, we need to cast the temperature dependent effective
potential in a convenient form.

The electroweak phase transition can be studied using a high-temperature
expansion of the effective potential, which, as shown by Turok and Zadrozny
\cite{Turok1} and Anderson and Hall \cite{Anderson and Hall}
is  very reliable in the relevent range of temperatures.
The $V^T[\phi]$ has two temperatures of interest
\begin{equation}
{T_{0}}^2 \ =\ \frac{1}{4D}({m_{H}}^{2} - 8B{v_{0}}^2)\qquad
{\rm and}\qquad
{T_{c}}^2 \ =\  \frac{{T_{o}}^2}{1-\frac{E^2}{{\lambda }D}}
\end{equation}
where, using the known Standard Model data, $D=0.04+0.06(m_t/125GeV)^2$,
$E\simeq0.01$, $v_0=246 GeV$ and $B\simeq-0.001(m_t/125GeV)^4$,
and we assume $\lambda \simeq \ m_H^2/2v_0^2$.
Details of the parameterisation can be found in \cite{Dine.et.al}.
$T_{0}$ is the temperature at which the symmetric phase becomes
an unstable extremum.  We rescale:
\begin{equation}
{\phi \ \longrightarrow \phi /  {(\frac{2E{T_c}}{\lambda })},\qquad
r\ \longrightarrow \ {(\frac{2E{T_c}}{\lambda })}\ r},\qquad {\rm and}\
{\bf t\ \longrightarrow \ (\frac{2E{T_c}}{\lambda})\ t}
\end{equation}
\noindent
Then  for any T such that $\mid {T_{c}-T}\mid \ \ll \ {T_{c}}$
the equation satisfied by $\phi$ is
\begin{equation}
{\frac{{\partial}^{2}{\phi}}{{\partial t}^{2}}} -
{\frac{{\partial}^{2}{\phi}}{{\partial r}^{2}}} -
 {{\frac{1}{r}}{\frac{\partial \phi}{\partial r}}} +{ \lambda{\phi}^3}
- {\frac{3}{2}}{\lambda}{\phi }^2
 + {\frac{\lambda}{2}}{(1 + \epsilon )\phi } \ = \ 0 \label{timdep}
\end{equation}
where,
\begin{equation}
\epsilon  \ = \ {\frac{D ( {T^2} - {T_0}^2) \lambda }{{E^2}{{T_c}^2}}}-1
 = \frac{T^{2}-{T_{c}}^2}{{T_{c}}^2 - {T_{0}}^2} \label{eps}
\end{equation}
For our calculation we take $m_t$ to be within the range $100 - 120 $ GeV.

We  used IMSL subroutine BVPFD for solving the time independent
equation. As initial trial, configurations with all $\phi_i = 0$ for grid
points with large $r$ were provided.
Solutions of class 1 were found to exist for $\epsilon \geq  - 0.07$.
For $\epsilon$ more negative BVPFD fails to return solutions with
the correct asymptotic behaviour, as supplied in the initial trial.
These configurations are defined unstable.  In such cases a solution is
 always found when a trial with all
$\phi_i = \phi_2$ for large $r$ is provided.

The value of $\ecr$ is mildly sensitive to $\lambda$\cite{Yajnik},
and hence $m_H$. For $m_H$ in the range 60 to 120 GeV, the critical
value varies between $-0.07$ and  $-0.09$.

Now to check the nature of time dependence of $\phi$, we solve the time
dependent equation (\ref{timdep})  using the IMSL subroutine MOLCH.
The initial data consisted of
the unstable configuration found for each $m_H$ by BVPFD and
initial time derivative zero. After intial period of slow evolution,
a wall interpolating between the two minima makes its appearance.
An example is shown in fig-1.

{}From above calculations we can obtain wall thickness and velocity. Wall
thickness is a parameter that is better understood. When the bubble radius
is large, the problem reduces to that of two dimensional soliton theory.
Whence thickness
\begin{equation}
\triangle r \ = \ \int_{\phi_{in}}^{\phi_{out}} \frac{d\phi }{\sqrt{2V^{T}(\phi
)}}
\end{equation}
where the limits of integration are the values of $\phi$ inside and
outside the bubble .
For definiteness we choose the limits $\phi_{in}\ =\ 0.75$ and
$\phi_{out}\ =\ 0.25$. These are the points of inflection of the scaled
$V_{\rm eff}$ curve. We find
\begin{equation}
\triangle r \ =\ s(m_H){T^{-1}}
\end{equation}
where $s(m_H)$ is a dimensionless scaling which we find to vary from
$0.7m_H$ /GeV to $0.5m_H$/GeV as $m_H$ changes from 60 to 120 GeV.  This
observation  is also borne out by the numerically
computed graphs of the bubble profile.

The correct estimate of wall velocity is more complicated. We can read it
off from a graph, superposing the wall profile at multiple instants of
time (fig-1). We find for the entire $m_H$ range 60 - 120 GeV, $v\ \sim
\ 0.5$. This is in accord with the relativistic detonation bubble wall
theory, as discussed by Steinhardt \cite{Steinhardt} and Turok
\cite{Turok2}  which predicts relativistic speeds with
no significant damping. Since the result is derived from
${V_{eff}}^{T}({\phi })$,
this result includes the nondissipative interactions with other particle
species. The dissipative interaction that may exist with the surrounding
plasma is also subleading in an expansion in v as demonstrated in
\cite{Turok2} and \cite{Dine.et.al}.
Analytical understanding of the wall velocity in our mechanism is very
difficult.  In the asymptotic region where the two
dimensional soliton theory applies, the forward acceleration of the wall
is
\begin{equation}
{\dot v}  =\
{\frac{3}{2}}{\frac{ET_{c}{\mid\epsilon\mid}}{\sqrt {2\lambda}}}
\end{equation}

where we have restored dimensions to $t$. This is a small accleration
since $E\epsilon \sim {10}^{-3}$.
Thus the large velocity is the initial condition on the soliton arising
from the event during which the bubble was formed. This happens in the
region where the ${\phi'} \ /{r}$ term is dominant, making analytic
estimates difficult. Any mild source of damping could help
the wall to reach a terminal velocity. The terminal velocity is expected
to be between 0.1 to 0.5.

$\ecr$ is a measure of the temperature $T_1$ at which the bubble can
be considered to have been formed, and the expansion of the wall to
have commenced. Using equation (\ref{eps}) we define $T_1$ by
\begin{equation}
\ecr  \equiv  \frac{T_1^{2}-{T_{c}}^2}{{T_{c}}^2 - {T_{0}}^2} \label{epscr}
\end{equation}
Typical numbers for $T_c$, $T_0$ and $T_1$ are $T_c=125$ GeV,
$T_c-T_0=1$ GeV and $T_c-T_1=0.08$ GeV.

Our general conclusion is that we have bubble formation promptly after
$T_c$ is reached and these bubbles have thick walls with relativistic
speeds. It can not be overemphasised that although we are considering
gauge extensions of the standard model, the bubble shape and velocity
are determined by weak scale physics. Similarly it can be seen from
equation (\ref{timdep}) that the value of ${\epsilon }_{\rm cr}$
depends if at all, on $\lambda$.
And we have checked that $\epsilon_{\rm cr}$ is insensitive to $\lambda$.
Thus the promptness of bubble formation
(smallness of $\mid {\epsilon_{\rm cr}} \mid$) is also universal .
The top quark mass enters parameters that are small compared to
those determined by the Higgs mass (in the range in which the
high temperature expansion used here is valid), and hence is unimportant
to the conclusions presented here.

We have shown that the parameters such as
bubble wall thickness and velocity are determined essentially by the
Standard Model physics. However the nature of the complete phase
transition depends on one additional feature. This is the length per unit
volume if the strings that can serve as the nucleation sites. The phase
transition begins by the nucleation of the true vacuum bubbles on the
strings. If there are sufficiently many such strings present, these
bubbles will expand, coalesce and complete the transition. If however the
strings are sparse, the transition has to be complemented by the formation
of spontaneous bubbles. Given that the strings formed at a temperature
$T\sim M_\chi$, there is no simple theory to trace their subsequent
evolution and the numerical computations are difficult. For this reason
definite predictions along these lines are impossible. However,
dimensional analysis combined with intuition regarding collective
phenomena\cite{Kibble} suggests that if $M_\chi$ is not greater than
$M_{ew}$ by more than a few orders of magnitude, there will be sufficiently
many strings present to complete the phase transition.

{}From the point of view of baryogenesis, the dynamics of the bubbles is very
important. For string induced bubbles to work for baryogenesis, we need a
mechanism that will work in thick, fast moving walls.
We have seen that the walls are thick $\sim 50$ to $100 {T_c}^{-1}$. These
provide adiabatic conditions for baryogenesis. In the criteria of
\cite{Cohenreview} we have
\begin{equation}
 {\tau}_{wall} \ = \ {\triangle r}\ /\ {v} \sim
10^{2} - {10^{3}}{T_c}^{-1}
\end{equation}
whereas equilibrating timescales for Standard
Model particles is ${\tau}_{T} \sim 10{T_c}^{-1}$.
Thus either the mechanism of McLerran Shaposhnikov Turok and Voloshin
\cite{MSTV} or of Cohen Kaplan Nelson \cite{Cohen1} can in principle work.
The general recipe of spontaneous baryogenesis of Cohen and Kaplan
\cite{Cohenreview},
\cite{Cohen2} is directly applicable here. We also see that extension
of the Higgs sector does not alter the main conclusions reached here
so long as at least one Higgs couples to the strings.
The extension of the Higgs sector is useful to electroweak
baryogenesis for two reasons. Firstly, it provides larger $CP$
violation, as for instance the  proposals of
\cite{MSTV} and \cite{Cohen1} which rely on the two Higgs doublet model.
Secondly, the extra Higgs aids the suppression of the washout
of baryon asymmetry in the broken symmetry phase.
It was shown by Anderson and Hall \cite{Anderson
and Hall} that simply adding a gauge singlet scalar with a biquadratic
coupling with the doublet Higgs substantially weakens the
upper bound on the Standard Model Higgs\cite{Bochka and Shapo}.
String induced phase transition easily accomodates such modifications.

In conclusion, the nature and dynamics of string induced
bubbles is determined by
electroweak physics. We have shown that the conditions in the walls are
adiabatic with regard to interaction time scales of the known particles.
For electroweak scale baryogenesis the Standard Model has to be extended to
have  {\bf B} or {\bf L} violation in
conjuction with sufficient {\bf CP} violation. Sphaleronic wash out in the
broken phase has to be prevented. The details of such extensions however
do not affect the conditions supplied by the walls for baryogenesis. This
reduces the uncertainties faced in constructing models of electroweak
baryogenesis.

One of us (S.B.D) wishes to thank Mr. D. Duari for his useful suggestions
about numerical methods.

\noindent
{\large{\bf Figure caption:}}\\
Figure-1 Time evolution of $\langle\phi\rangle$
\vfil



\end{document}